# X-BAND DEFLECTING CAVITY DESIGN FOR ULTRA-SHORT BUNCH LENGTH MEASUREMENT OF SXFEL AT SINAP[1]

TAN Jian-Hao[1,3](谭建豪) GU Qiang[1] (顾强) FANG Wen-Cheng[1] (方文程)TONG De-Chun[2] (童德春) Zhao Zhen-Tang[2](赵振堂)C

[1] Shanghai Institute of Applied Physics, Chinese Academy of Science, Shanghai 201800, China;

[2] Department of Engineering Physics, Tsinghua University, Beijing 100084, China;

[3] Graduate University of Chinese Academy of Science, Beijing 100039, China

[4] Shanghai Key Laboratory of Cryogenics＆Superconducting RF Technology, Shanghai 201800, China

Abstract:

For the development of the X-ray Free Electron Lasers test facility (SXFEL) at SINAP, ultra-short bunch is the crucial requirement for excellent lasing performance. It's a big challenge for deflecting cavity to measure the length of ultra-short bunch, and higher deflecting gradient is required for higher measurement resolution. X-band travelling wave deflecting structure has features of higher deflecting voltage and compact structure, which is good performance at ultra-short bunch length measurement. In this paper, a new X-band deflecting structure has been designed operated at HEM11-2pi/3 mode. For suppressing the polarization of deflecting plane of the HEM11 mode, two symmetrical caves are added on the cavity wall to separate two polarized modes. More details of design and simulation results are presented in this paper.

Key words:HEM11, bunch length, dipole mode, polarization, time domain.

1 Introduction

The 840MeV linac based X-ray free electron lasers test facility will be built at the Zhangjiang campus of shanghai institute of apply physics (SINAP), where Shanghai Synchrotron Radiation Facility (SSRF) had been constructed [1]. SXFEL will be a compact coherent X-ray source, and its total length is about 300 meters. This facility will generate electron beams with low emittance and short bunch length, which is about 2.5um and 120um, respectively. For ultra-short electron bunch length measurement, a standard method has been used at LCLS with a new X-band deflecting structure, and the principle of bunch length measurement works as well as stream camera for electrons, however, the deflector is capable of resolving bunch length as short as 10 femtosecond. Deflecting cavity operates in a dipole mode, for suppressing the polarization of deflection plane of the HEM11 mode, with two symmetrical caves added on the cavity wall [2]. In order to diagnose longitudinal structure of the bunch, a novel transverse RF deflecting structure is introduced as shown in figure 1.

[1] Supported by Knowledge Innovation Project of The Chinese Academy of Science (455011061)

* email: zhaozhentang@sinap.ac.cn)

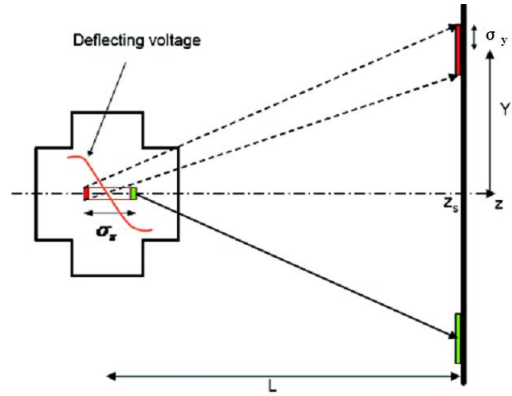

Figure 1 Schematic of an RF deflector

2 Demanded performance

In SXFEL, the deflecting structure will locate after the last bunch compressor within 3 m at a beam energy of 0.84GeV. As mentioned above, the emittance and bunch length are 2.5um and 120um, respectively. The relation between deflecting voltage $V_T$ and beam parameters is given by

$$eV_T = \frac{E\lambda/c}{2\pi\sigma_t R_{34}}\sqrt{\sigma_y^2 - \sigma_{y,0}^2} \quad \text{(1)}$$

Where $V_T$ is the transverse deflecting voltage, and E is the electron energy, $\sigma_t$ is the longitudinal beam size on the time scale, $\lambda$ is the RF wavelength, $R_{34}$ is the element of the transport matrix, $\sigma_y$ and $\sigma_{y,0}$ is the vertical beam size on the screen when the deflector is on and off. When a length of drift space between the deflector and screen is taken into consideration, $R_{34}$ is the drift space length D. When there is a complicated system, such as a focusing system composed of several quadrupoles, $R_{34}$ can be acquired by multiplying these transport matrices. Further general condition, $R_{34}$ could be expressed as $R_{34} = \sqrt{\beta_d \beta_s}\sin\psi$, where $\beta_d$ and $\beta_s$ are the beta function at the deflector and screen respectively, $\psi$ is the phase shift between the beta oscillation at the deflector and screen on deflection plane. For deflecting structure, the resolution is a very important conception which defined as $\Delta_t = \frac{E}{\omega}\frac{\sigma_{y,0}}{V_{def}R_{34}}$, where $\sigma_{y,0}$ can be stated by $\sqrt{\frac{\beta_s \varepsilon_{N,y}}{\gamma}}$ [3]. Higher time resolution can be acquired in the condition of higher frequency and voltage, lower energy and emittance, larger drift space or beta function at the deflector are also preferred. Considering the costs, a lower power-feed structure program is proposed due to enough space. In addition, for the development of compact facilities, smaller, shorter, and higher gradient structure is the main target for structure, therefore, an x-band disk-loaded waveguide structure is the best selection of the design.

Table 1 Specifications of x-band deflector for SXFEL

| Structure type | Constant impedance |
|---|---|
| Operating frequency | 11.424GHz |

| Operating mode | Disk-loaded waveguide 2pi/3 |
|---|---|
| Energy | 0.84GeV |
| Bunch length | 120um |
| Bunch size(y) | 60um |
| Drift space | 2-3m |
| Resolution | 20-10fs |
| Deflecting voltage | 12-35MV |
| Input power | 20 MW |
| total length L | 1m |

Table 1 shows the specifications of x-band deflector for SXFEL, which will working at traveling wave state, HEM11 mode. The performance of deflector is determined by transverse shut impedance, group velocity and attenuation factor, Eq.2 gives the relationship between these parameters.

$$V_T^2 = P_0(1-e^{-2\tau})R_s L \quad\quad\quad (2)$$

3 Design of regular cells

The HEM11 mode in an axis-symmetric structure degenerates in twofold, and it leads to the rotation of deflection plane or polarization plane which brings some inconvenience when deflecting cavity is used to measuring the beam size. To solve the degeneracy and prevent rotation of the deflection plane, SLAC has proposed design project with two symmetrical holes on the iris which called LOLA in 1960s [4], and another type with two symmetric caves added on the cavity wall [5], for cancelling the degeneracy of the HEM11 mode, an racetrack structure design as well at Spring-8 [6], then the frequency of the degenerate modes exist mode gap which depend on the degree of the symmetries-broken.

3.1 Selection of scheme

To make the fabrication process easier, the influence on frequency by the added variation are simulated. Figure 2 shows the impact of different types design.

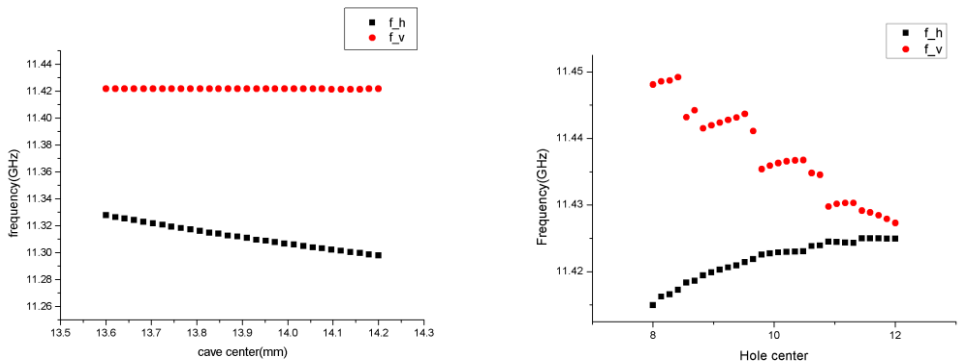

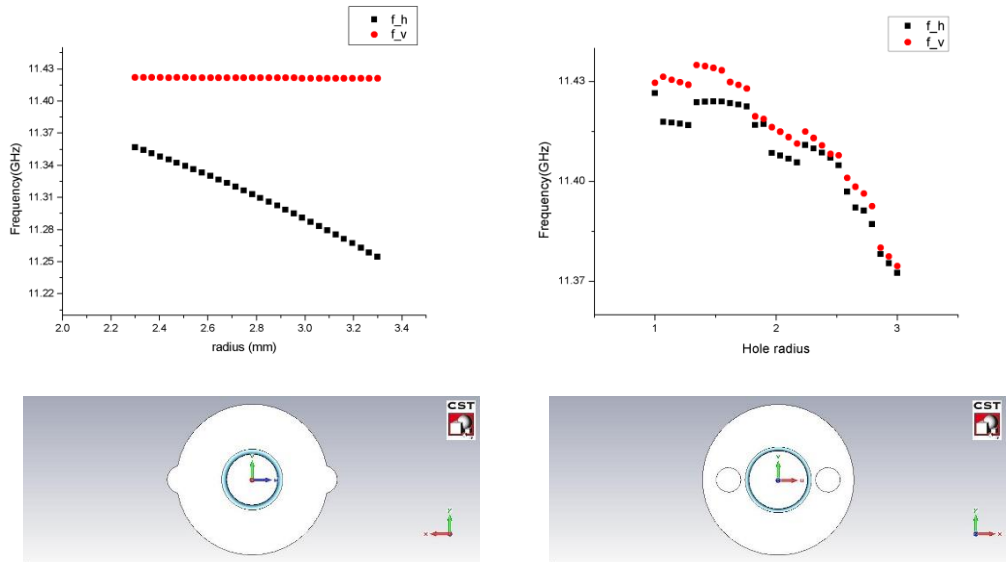

Figure 2 Frequency affected by added variation

It is obvious that, the caves type structure is superior to holes type structure. The position of the caves center and the radius of the caves almost have no effect on perpendicular direction, but have a significant influence on horizontal direction (here caves or holes added on horizontal direction). Hence, the SLAC caves type structure are applied on the deflector design. HEM11 mode, different from accelerating mode, its dispersion diagrams will change from forward to backward when a/b vary from large to small [7]. Besides, the dispersion characteristic of HEM11 mode is a hybrid of TM11 and TE11 mode, hence, the single chain of coupled circuit model is not suitable for HEM11, instead of double chain of circuit model which means the field in each cell is expand into a combination of a TM11 and a TE11 mode. The details are presented in Ref [8], and the final expression is Eq. 3

$$\cos\theta = \frac{-k_1 k_2 - (1-\frac{f_1^2}{f^2})(1-\frac{f_2^2}{f^2})}{(1-\frac{f_1^2}{f^2})k_2 + (1-\frac{f_2^2}{f^2})k_1} \quad \text{.............................................. (3)}$$

Where k, f are the couple coefficient and regular frequency, respectively, the index 1 and 2 represent TM11 and TE11 mode, respectively. The results with double chain of circuit model had been calculated, and single chain of circuit model which is used to compare with the former also applied to the calculation. The results of double chain of circuits model are consistent with the dispersion curve of HEM11 mode which simulated by CST as shown in the figure 3.

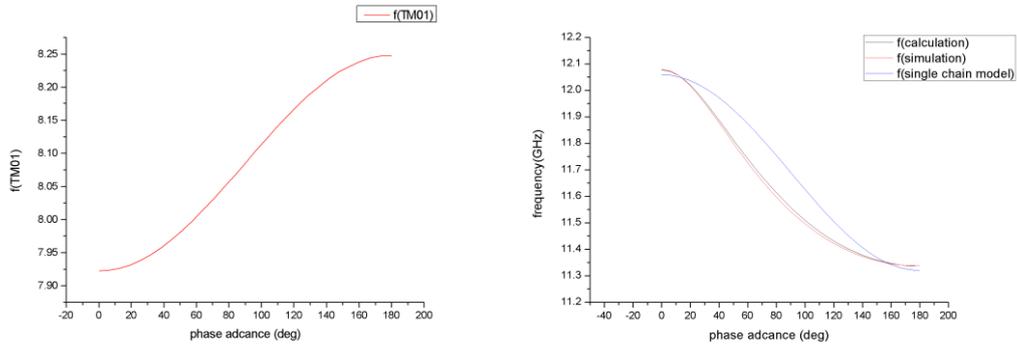

Figure 3 Single chain and double chain of circuit model results

The definition of group velocity and attenuation factor are $V_g = \dfrac{d\omega}{d\beta}$ and $\alpha = \dfrac{\pi f}{QV_g}$, take advantage of the dispersion curve, it is easy to calculate the group velocity, and the attenuation factor could be fitted when the quality factor (Q) of the cell is clear. It is fortunate that the Q value is given by the template post processing of CST. The group velocity indicate the velocity of power flow which determines the scale of feed-in power, and high power flow represent the high power accumulation which would leads to high breakdown rates. Therefore, the cavity with low group velocity is a more appropriate choice than high group velocity for high gradient structure. To utilizing the power maximum, the cells with high shut impedance are preferred. The longitudinal shut impedance is defined as

$$R = \dfrac{E_0^2}{P} \quad \text{............................................................................ (4)}$$

It is an important performance for accelerating structure, but for deflecting structure, the transverse shunt impedance is more useful. The transverse shunt impedance $R_s$ is defined as [9]:

$$R = \dfrac{[\dfrac{1}{k}\dfrac{\partial E_z}{\partial r}]^2}{-\dfrac{dP}{dz}} \quad \text{.............................................................. (5)}$$

Where z and r is longitudinal and transverse axis respectively, $E_z$ is the electric field amplitude for the dipole mode with angular frequency $\omega$, and P is the RF power as function of z. The expression of shut impedance shows that the structure with high shut impedance would get higher gradient with the same level input power. With the assist of the data from single cell simulation results, transverse shut impedance could be calculated by Eq.(4), each different r0 corresponds to different $R_s$, then it gets a fitting curve, which shows the transverse shut impedance on each point and gets the mean value of the transverse shut impedance of the cavity. As shown in figure4.

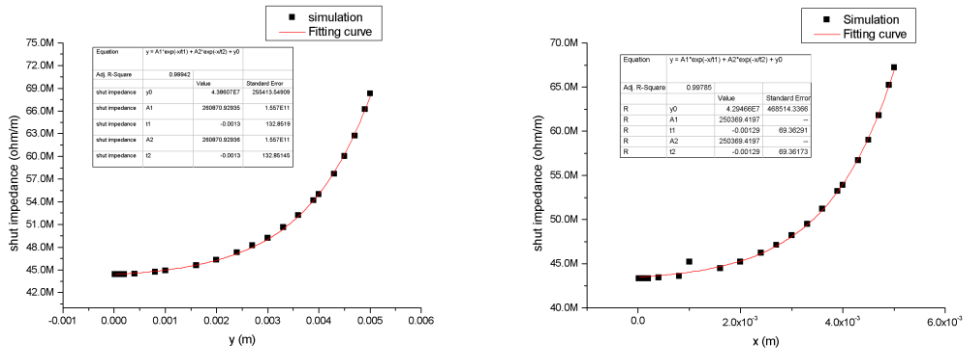

Figure 4 Transverse shut impedance

3.2 Simulation of regular cells

Now that the caves type has been selected to apply on X-band deflector design, and the required performance has been proposed, then the regular cells are simulated and several key parameters which evaluate the property of cells are calculated. As mentioned above, the ratio a/b has an impact on power flow direction, when a/b varies from large to small, the group velocity changes from positive to negative, which means the power flow direction and beam direction are of opposite direction. Furthermore, the scale of 2a has an effect on transverse shut impedance, group velocity, and attenuation factor. Therefore, a comparison among transverse shut impedance, group velocity and attenuation factor of different a/b with frequency fixed at 11.424GHz.

Table 2 Comparison for HEM11 mode as a function of a/b

| a (a/b) (mm) | Group velocity (%c) | Transverse shut impedance (MΩ/m) | Quality factor | Attenuation factor (1/m) |
|---|---|---|---|---|
| 5.5(0.37) | -2.46 | 41.120 | 6662 | 0.730 |
| 5.2(0.35) | -2.98 | 45.119 | 6622 | 0.606 |
| 5.0(0.34) | -3.17 | 49.072 | 6778 | 0.564 |
| 4.5(0.30) | -3.18 | 56.780 | 6924 | 0.543 |

Table 2 gives the summary of group velocity, transverse shut impedance, attenuation factor as a function of a/b. As the ratio a/b decrease, the group velocity, transverse shut impedance and quality factor have an increasing trend, however, the attenuation factor is on the decline. Considering the room for deflector is enough, high power operation is not the main purpose, in addition, lower group velocity leads to stronger dispersion. Therefore, it is an optimum scheme that the radius of beam hole is selected as 4.5mm.

Considering the strong electric field on the iris, as shown in figure 5, a comparison for

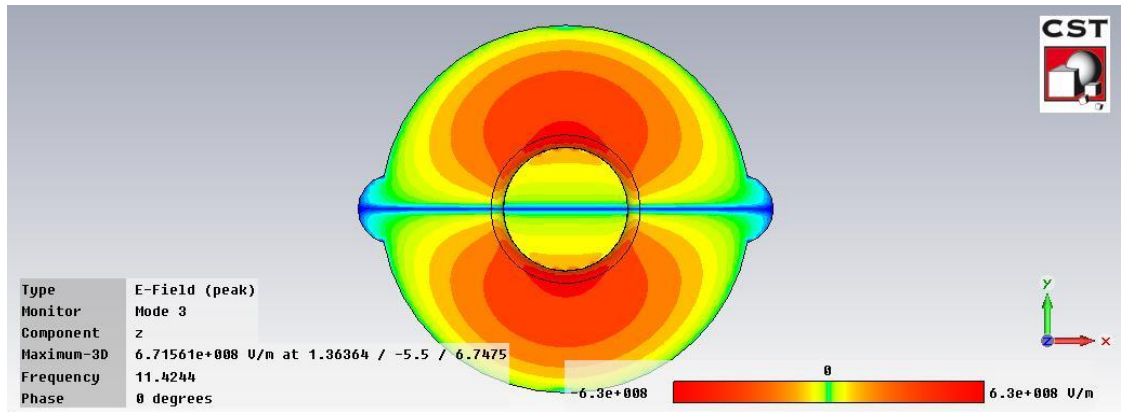

Figure 5 electric field distribution in the regular cell

HEM11 mode as iris thickness t are simulated. Table 3 gives the summary of each specification vary by t, the group velocity almost at the same, when t vary from large to small, the transverse shut impedance and quality factor have a little increase, on the contrary, the attenuation factor have a small decrease. Considering such an improvement on the performance of the cell, and the degree of improvement, the thickness of iris selected as 1.8mm is a reasonable choice.

Table 3 Comparison for HEM11 mode as a function of t

| t (mm) | Transverse shut impedance(MΩ/m) | Group velocity(%c) | Q value | Attenuation factor(1/m) |
|---|---|---|---|---|
| 2.2 | 45.62 | 3.17 | 6345 | 0.595 |
| 2.0 | 49.07 | 3.17 | 6778 | 0.557 |
| 1.8 | 50.31 | 3.16 | 6965 | 0.544 |

Eventually, the parameters of the cells have been selected, based on the available beam parameters and the performance of the cells, the feed-power and the length of the structure can be calculated. The suitable length of the deflector could be estimated with the Eq. 2. About a 1 meter long deflector was selected when it will be utilized to measuring the bunch length, where have taken the power utilization efficiency, the space to install the deflector and the input power level to consideration. Therefore, more than 100 regular cells are needed [10]. Thus, the final specification of the regular cells are listed in table 4.

Table 4 Initial scheme of deflectors for 20-fs temporal resolution

| Structure type | Constant impedance |
|---|---|
| Operating frequency | 11.424GHz |
| Operating mode | Disk-loaded waveguide 2pi/3 |
| Total length L | 1 m |
| Resolution | 20fs |
| Deflecting voltage | 11 MV |
| Input power | 5 MW |
| Group velocity Vg | 3.17%c |
| Filling time $t_F$ | 98 ns |

4 Design of coupler

The regular cells had been designed and provide a plenty of outstanding performance, however, it is very important to feed the power into the cavity and maintain a high efficiency. The RF coupler is an important element for microwave structure. The primary mission of couplers is feed the power from klystron to the regular cells as much as possible. For standing wave structure, a feed port is enough owing to the characteristic of standing wave, whereas two couplers are needed in travelling wave structure. All the regular cells have the same size because of constant impedance structures; hence the time domain could be used for simulation of coupler [11].

Several different types of coupler, including mode launcher coupler, waveguide coupler and standard coupler, as shown in Fig.6, which one is more suitable for deflecting cavity on earth.

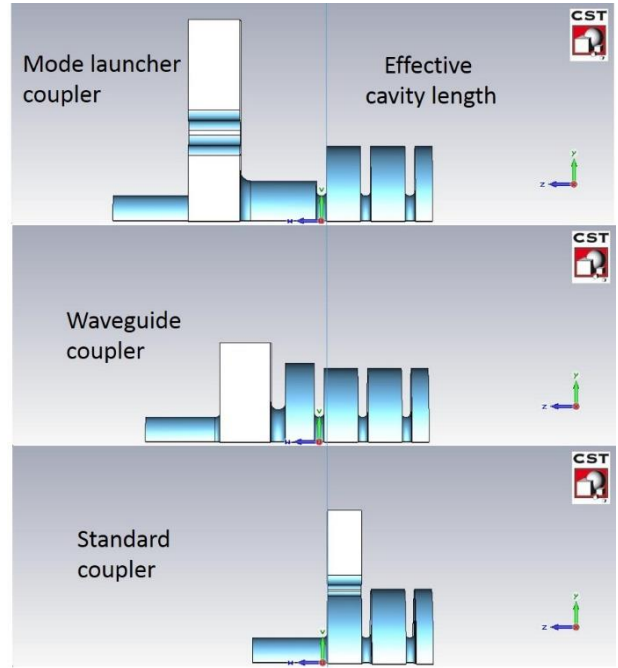

Figure 6 Three types of couplers

As a matter of fact, mode launcher coupler and waveguide coupler are dual-feed coupler generally. Comparing with single-feed coupler, the improvement of multipole field components is negligible [12], and the dipole components are in all cells with single-feed coupler. Take the total length and the efficiency to consideration, the standard coupler is more meeting the request than others. Single-feed standard coupler is the optimized choice, as shown in figure 7 finally.

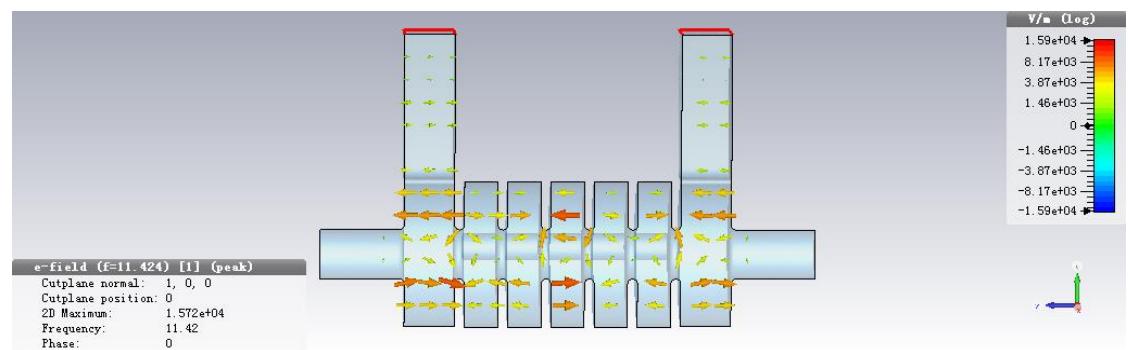

Figure 7 Single-feed standard coupler models and field distribution

The simulation was set with five identical regular cells and two identical (input/output) couplers owing to constant impedance structure. The results of simulation and calculation are illustrated in figure 8 and figure 9.

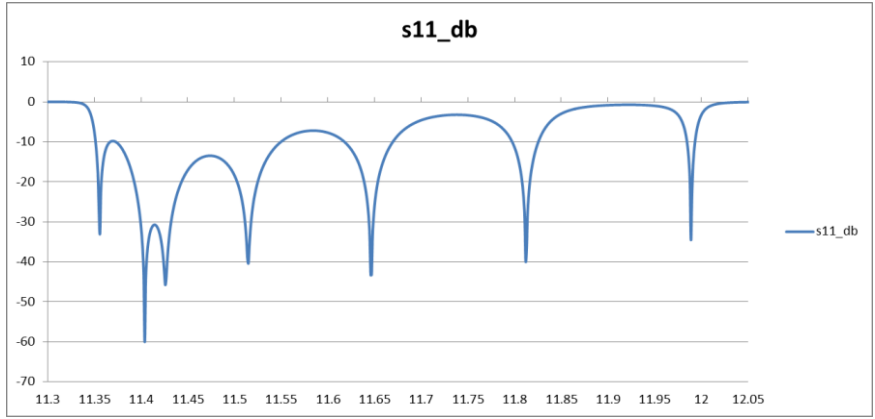

Figure 8 $S_{11}$ parameter

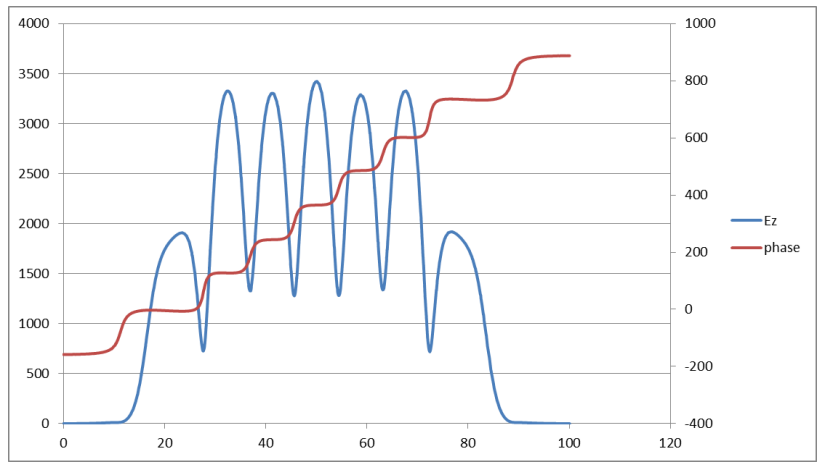

Figure 9 Z component of electric field and phase distribution

Due to the electric field along z axis is null, the off-axis electric field has been taken to consideration. The simulation results (Fig.9) present that the field in the coupler is slightly lower than the cavity, and shows the periodic characteristic of structures too, the above results seem like the couplers have been matched. It is well known that, the tuning of the structure including two parts which are the cavity and the couplers, respectively. It is significant to ensure the phase shift of the cavity is corresponding to the operating mode and the reflection coefficient of the coupler is lower than 0.03 to eliminate the standing wave in the cavity. Time domain simulation methods, the Eq.7 to Eq.9 below, could check the couplers match or not.

$$F^{\pm}(z) = \frac{E_c(z+p)}{E_c(z)} \quad \quad (7)$$

$$\Sigma(z) = F^{+}(z) + F^{-}(z)$$
$$\Delta(z) = F^{+}(z) - F^{-}(z) \quad \quad (8)$$

$$\cos\phi = \frac{1}{2}\Sigma(z)$$
$$Re^{2j\Phi} = \frac{2\sin\phi + j\Delta(z)}{2\sin\phi - j\Delta(z)} \quad \quad (9)$$

Where E(z), $\phi$, R are the distribution of electric field along z, regular cell phase advance and reflection coefficient, respectively.

Figure.10 presents the calculation result, and all results demonstrate that the couplers are matched and the phase advance is 120 deg. It is obvious that the two values $\cos\phi$ and R are indeed constant over the allowed range, and $\phi$ is corresponding to phase shift 2pi/3. It suggests that the regular cells operate at the frequency we wanted. The reflection coefficient R demonstrates that the couplers have satisfied the matching condition.

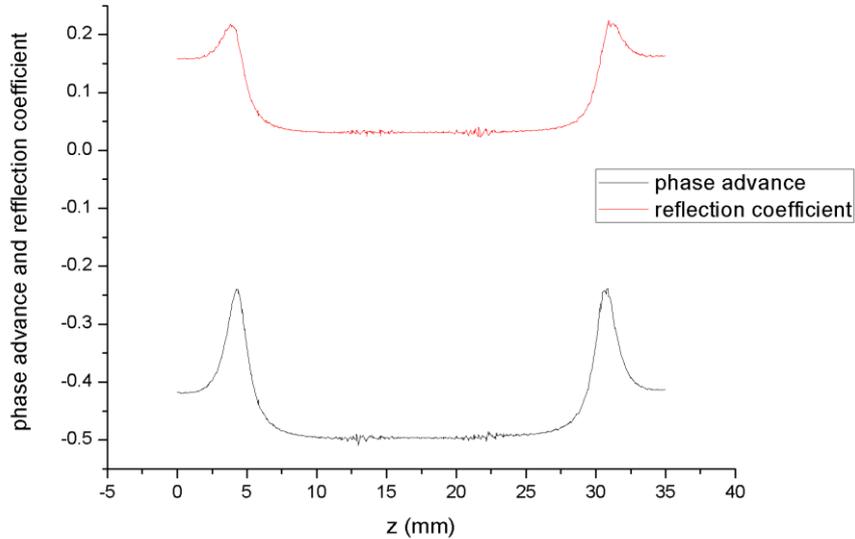

Figure 10 Reflection coefficient and phase advance (matched)

4 Conclusions

A transverse travelling wave deflecting structure of a disk-loaded waveguide has been designed and optimized for SXFEL at SINAP, with two caves in horizontal orientation to stable the polarization plane. The transverse shunt impedance could be up to $50\,M\Omega/m$, therefore a 110-cell long deflecting structure fed with an RF power of 5 MW generate a total deflecting voltage of 11 MV, then we can measure the bunch length with a time resolution of 20fs by using this high gradient deflector. For higher energy or getting higher resolution, improving the fed power could meet the requirement. In the next step, a prototype of the X-band deflecting cavity will be manufactured, and its beam test will be carried out at SXFEL in the near future.

Acknowledgement

We are grateful to Meng Zhang and Dazhang Huang for valuable discussions and suggestions. We also would like to thank many other staff members at the LINAC group of FEL department SINAP for their supports with ideals and discussions.

REFERENCES
[1] FENG Chao, ZHAO Zhen-Tang et al. Hard X-ray free-electron laser based on echo-enabled staged harmonic generation scheme. Chinese Sci Bull, 2010, 55: 221-227
[2] WANG Ju-Wen, TANTAWI S et al. X-band travelling wave RF deflector structure. Proceedings of LINAC08, Victoria, BC, Canada

[3] Paul Emma al. A Transverse RF Deflecting Structure for Bunch Length and Phase Space Diagnostics, LCLS-TN-00-12

[4] AKRE R et al. SLAC, Stanford, CA 94309, USA. Bunch length measurement using a transverse RF deflecting structure in the SLAC linac. Proceedings of EPAC 2002, Paris, France.

[5] WANG Ju-Wen. X-band Deflectors development at SLAC. December 2008.

[6] Hiroyasu Ego et al. Design of the transverse c-band deflecting structure for measurement of bunch length in X-FEL. Proceeding of EPAC08, Genoa, Italy

[7] WANG Ju-Wen et al. HEM11 modes revisited. SLAC-PUB-5321 September 1990

[8] Bane K L F et al. The coupled dipole modes of the NLC accelerator structure. SLAC-PUB-5766 March 1992(A)

[9] LOEW G A et al. The definition and measurement of the shut impedance of an rf particle deflector. Slac-tn-62-057 1962

[10] ZHAO Zhen-Tang et al. SINAP. Shanghai soft x-ray free electron laser test facility. Proceedings of IPAC2011, San Sebastian, Spain

[11] KROLL N M et al, SLAC. Applications of time domain simulation to coupler design for periodic structure. Proceeding of LINAC2000, Monterey, CA, USA

[12] AMBATTU P K et al, Cockcroft Institute/Lancaster University. Coupler induced monopole component and its minimization in deflecting cavities. PHYSICAL REVIEW SPECIAL TOPICS, ACCELERATORS AND BEAMS 16, 062001(2013)